\documentclass[aps,prl,twocolumn,english,showpacs,letterpaper]{revtex4-1}

\usepackage[dvips]{epsfig}
\usepackage[colorlinks=true,citecolor=blue,linkcolor=blue]{hyperref}
\usepackage{amssymb}
\usepackage{graphicx}
\usepackage{subfigure}
\usepackage{array}

\usepackage{epsfig}
\usepackage{amsmath}
\usepackage{xcolor}
\usepackage{float}
\usepackage{mathrsfs}
\usepackage{indentfirst}
\usepackage{textcomp}
\usepackage{mathtools}

\newcommand{\red}[1]{\textcolor{red}{#1}}

\newcommand{\ignore}[1]{}

\newcommand{\aortho}{{\textbf{a}$_{\mathrm{ortho}}$}}
\newcommand{\bortho}{{\textbf{b}$_{\mathrm{ortho}}$}}
\newcommand{\cortho}{{\textbf{c}$_{\mathrm{ortho}}$}}
\newcommand{\Pn}{\textit{Pn}}

\newcommand{\COMMENTED}[1]{}

\newcommand{\titledef}{On the origin of metal-insulator transitions in the parent compounds of ruthenium-pnictide superconductors}

\begin{document}

\title{\titledef}

\author{Niraj Aryal}
\email{naryal@bnl.gov}
\affiliation{Condensed Matter Physics and Materials Science Division, Brookhaven National Laboratory, Upton, New York 11973, USA}
\author{Emil S. Bozin}
\affiliation{Condensed Matter Physics and Materials Science Division, Brookhaven National Laboratory, Upton, New York 11973, USA}
\author{Weiguo Yin}
\email{wyin@bnl.gov}
\affiliation{Condensed Matter Physics and Materials Science Division, Brookhaven National Laboratory, Upton, New York 11973, USA}

\date{\today}

\begin{abstract}
We study the interplay of the structural phase transition, flat electronic band dispersion, and metal-to-insulator transition (MIT) in the parent compounds of the Ru-pnictide superconductors by using first-principles calculations. 
Our electron and phonon calculations reveal that Ru(P,As) undergo MIT accompanied by orthorhombic to monoclinic distortion at low temperature, but RuSb stays orthorhombic and metallic in agreement with the experimental findings. 
We find that although small monoclinic distortion can remove the van Hove singularity at the Fermi level, it does not immediately  gap out the Fermi surface and a  large value of monoclinic distortion is necessary for a clear MIT suggesting the possibility of an intermediate pseudogapped monoclinic metallic phase.
Furthermore, we predict a light-induced two-step insulator-to-metal and structural transitions in the monoclinic phases of RuP and RuAs, which can be tested in future ultrafast pump-probe experiments as an alternative ideal play ground to VO$_2$.


\end{abstract}

\maketitle

\emph{Introduction.}---Understanding the origin of superconductivity and its interplay with charge-density-wave (CDW) instability, band topology, and metal-to-insulator transition (MIT) continues to be an important problem in condensed matter physics~\cite{HighTc_Fradkin_RMP2015,TiSe2_Kusmartseva_PRL2009,Kagome_Yu_NatComm2021}.
The discoveries of high-temperature superconductivity and many competing phases in layered copper oxides~\cite{Bednorz_ZPB_86_BaLaCuO,Wu_PRL_87_YBCO} and iron pnictides (Fe\Pn)~\cite{FeAs_JACS2008,LnFeAs_Lee_JPSJ2008,SmFeAs_Chen_Nature2008,CeFeAs_Chen_PRl2008} have triggered numerous investigations to find superconductivity in similar materials such as layered nickelates~\cite{Li_Nature_19_NdNiO2,Pan_NM_22_Nd6Ni5O12,Sun_Nature_23_La3Ni2O7} and pnictides~\cite{CrAs_Wu_NatCom2015,CrAs_Kotegawa_JPSJ2014,MnP_Cheng_PRL2015} for possible hints on the mechanism of high-temperature superconductivity and its relationship with the interplay of the charge, spin, orbital, and lattice degrees of freedom. To discern this complex interplay via light-induced transient phase transition is of particular interest due to recent advancements in ultrafast measurements~\cite{Fausti_Science_11,Mankowsky_Nature_14_YBCO,Nevola_PRX_23_FeTeSe}; however, it turns out to be highly challenging to even understand the light-induced MIT and melting of CDW in VO$_2$, especially concerning the emergence of the instantaneous monoclinic metallic phase~\cite{Xu_NC_23_VO2_UED,Li_PRX_22_VO2_UED,Wall_Science_18_VO2_ultrafast-XRD,Morrison_Science_14_VO2_UED,Wegkamp_PRL_14_VO2_ultrafast-ARPES,Baum_Science_07_VO2_UED}. Here, we show that the recently discovered ruthenium-pnictide (Ru\Pn) family of superconductors Ru$_{1-x}$Rh$_x$As, Ru$_{1-x}$Rh$_x$P, and RuSb near a CDW-driven MIT~\cite{MIT_Harai_PRB2012,OpticsRuP_Chen_PRB2015,NMR_RuP_FanCPL2015,NMR_RhRuP_Li_PRB2017,AnisotropicMIT_NakajimaPRB2019,3c4eRuP_Hirai_JACS2022} provides an alternative ideal play ground.

Ru{\Pn} (\Pn=P, As, Sb) materials crystallize in the MnP-type orthorhombic structure with space group of \textit{Pnma} (no. 62) at high temperature.
Upon cooling, two successive MITs have been reported for Ru(P,As). At the first transition temperature $T_1$, these two materials undergo metal to pseudo-gap (PG) phase transition and at $T_2$, there is a transition to the non-magnetic monoclinic insulating phase.
The presence of the PG feature and dome-shaped superconductivity upon doping is reminiscent of the behavior of Fe\Pn~\cite{PGFeAs_Ishida_JPSJ2008,PGFeAs_Sato_JPSJ2008,PGFeAs_Shimojima_PRB2014} and cuprate superconductors~\cite{Warren_PRL1989,NMR_PG_Alloul_PRL1989,PGReview_NormanPines_AIP2005}.
In addition, recently Koch {\sl et al.} reported the observation of fluctuating local Ru-trimers well above $T_1$ in synchrotron x-ray total scattering measurements~\cite{RuPEmil2022}. This finding, in contrast to VO$_2$ where no such evidence of high-temperature fluctuating local V-dimers are present~\cite{Corr_PRL_10_VO2_PDF}, might facilitate the emergence of the much debated instantaneous monoclinic metallic phase in pump-probe experiments as well as could shed light on the PG behaviour.
RuSb, on the other hand, stays metallic with no structural phase transition and is superconducting without doping.
 Unexpectedly, experiments have indicated higher density of states at the Fermi level in RuSb; hence the finding of monoclincity in RuP and RuAs appears to contradict the common notion that higher DOS drives instability~\cite{MIT_Harai_PRB2012}. Therefore, it is important to thoroughly investigate the electronic structure, the driving mechanism for the CDW transition and its relation to the MIT in these family of materials.
As Ru lies one row below Fe in the periodic table, Ru\Pn~ and Fe\Pn~ could share similar properties; one the other hand, the $4d$ valence electronic states in Ru\Pn~ are expected to be less localized and less magnetic than the $3d$ states in Fe\Pn, making Ru\Pn~ more tractable theoretically.

In this letter, we perform a systematic study of the electronic bands, phonon bands, and electron-lattice coupling in Ru(P,As,Sb) materials using first-principles calculations in order to understand the origin of metal-to-insulator transition. We find that the high temperature orthorhombic phase of these systems hosts symmetry-enforced multiple Dirac band crossings and multi-fold band degeneracies in the vicinity of the Fermi level.
For RuP and RuAs, the symmetry-enforced degenerate flat bands are present right at the Fermi level giving rise to the van Hove singularity whereas the bands are less flat and shifted away from the Fermi level for RuSb.
The monoclinic distorted phases in Ru(P,As) are energetically favorable compared to the orthorhombic phase whereas RuSb prefers orthorhombicity which is consistent with experiments.
Furthermore, we predict a light-induced two-step insulator-to-metal and structural transitions in the monoclinic phases of RuP and RuAs---with an intermediate monoclinic metallic phase---which can be attested in future ultrafast pump-probe experiments as an alternative ideal play ground to VO$_2$.

\begin{figure}[t]
    \begin{center}
            \includegraphics[width=0.48\textwidth]{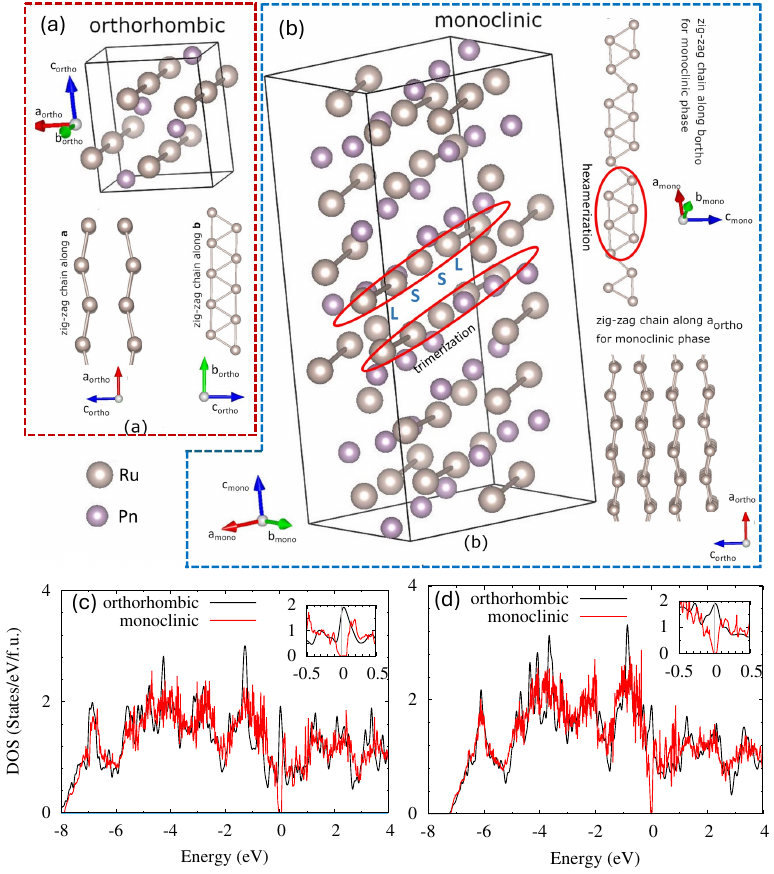}
            \end{center}
            \caption{(a) and (b) Crystal structure of the Ru-pnictides for orthorhombic and monoclinic phases, respectively, shown from different perspectives (see text for details). 
            (c) and (d) Density of states plot for RuP and RuAs showing comparison between the orthorhombic and monoclinic phases. The inset plots show DOS in $\pm$0.5 eV window around the Fermi level.        }
        \label{fig:unit cell} 
\end{figure}

\begin{figure*}[t]
    \begin{center}
            \includegraphics[width=0.98\textwidth]{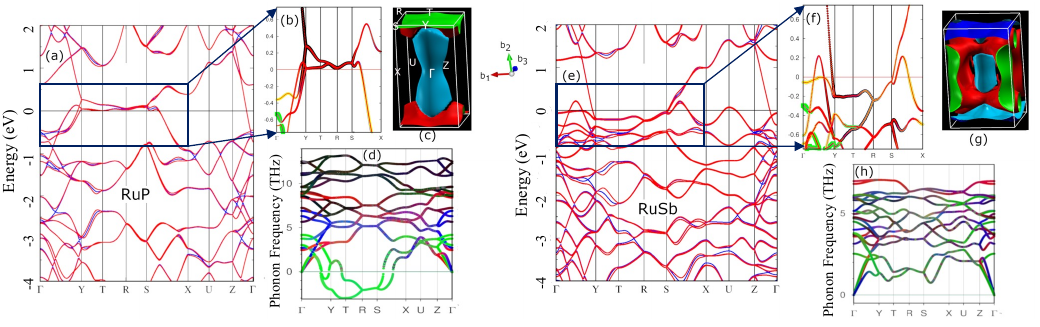}
            \end{center}
            \caption{Comparison of the electronic band structure, Fermi surface and phonon band dispersion between the orthorhombic phases of RuP and RuSb. Red (blue) lines in panels (a) and (e)  are obtained with (without) the inclusion of SOC. Panels (b) and (f) show Ru-d band character plots in a small window around the flat bands with the following
            color convention: black ($d_{xy}$), red ($d_{z^2}$), green ($d_{x^2-y^2}$), blue ($d_{xz}$), and yellow ($d_{yz}$).
            (c) and (f) show Fermi surface plots. The red-green (blue-magenta) colors show the hole (electron) Fermi surface pieces.
            (d) and (h) show the phonon band dispersion. The colors denote the weight of the normal mode atomic displacements along the crystallographic \aortho (red), \bortho (green) and \cortho (blue) direction for Ru-atoms. Black color shows the weight of pnictogen atoms.
        }
        \label{fig:fig2} 
\end{figure*}

\emph{Crystal structure and lattice relaxation.}---Fig.~\ref{fig:unit cell}(a) and (b) shows the crystal structure of Ru-pnictides in both orthorhombic and monoclinic distorted phase.
Above $T_1$, the system crystallizes in an orthorhombic  \textit{Pnma} space group with 4 formula units in an unit cell.
The Ru-atoms are surrounded by 6 \Pn-atoms in a distorted octahedral environment with face-sharing octahedra along the \aortho-direction and edge-sharing ones along the \bortho-direction and \textbf{bc}-plane.
A zig-zag chain of nearest neighbour (nn) Ru-atoms runs along the \textbf{a}-direction.
In addition, the next nn and third nn Ru-atoms form a triangular lattice in the \textbf{bc}-plane.
Below $T_2$, P and As members of the family undergo orthorhombic to monoclinic distortion.
The unit cell triples both along the orthorhombic \textbf{c}-direction and in the \textbf{ab}-plane thereby increasing the number of atoms in the monoclinic unit cell by 9 times.
The transformation matrix that maps from orthorhombic to monoclinic unit cell is given by $\mathrm{V_{mono}} = \mathcal{M} \mathrm{V_{ortho}}$
where, $\mathrm{V}$'s are the lattice vectors for the monoclinic and orthorhombic unit cells~ncite{CommentCrystalConvention} and $\mathcal{M}= 
\begin{bmatrix}
1 & 2 & 0\\
-1 & 1 & 0\\
0 & 0 & 3\\
\end{bmatrix}.$ 
    The monoclinic distorted phase is characterized by trimerization along the \bortho-direction consisting of two short (S$_1$, S$_2$) and one long (L) bonds in the order ---LS$_1$S$_2$LS$_1$S$_2$L---.
Such trimerization results in tightly bound Ru$_6$ clusters or hexamers. In addition to the trimerization and hexamerization, the zig-zag chain along \aortho direction has a weak modulation of the bond distances with a period of 6.

Fig.~\ref{fig:unit cell} (c) and (d) shows the density of states comparison between the orthorhombic and monoclinic distorted phase for both RuP and RuAs obtained from Density Functional Theory (DFT) calculation. The orthorhombic phases show peak DOS at the Fermi level thereby signalling the presence of lattice instability. 
As we will see below, such peak DOS arises from the presence of Ru-$d_{xy}$ derived flat bands at the Fermi level.
The fully relaxed monoclinic distorted phases correctly reproduces the insulating phase for both systems with gap-size comparable to the experimental findings which is $\sim100$~meV for RuP~\cite{NMR_RhRuP_Li_PRB2017,RuP_MIT_Ootsuki_PRB2020,RuPEmil2022}, in agreement with a recent DFT study of RuP~\cite{3c4eRuP_Hirai_JACS2022}, and $\sim30$~meV for RuAs~\cite{RuAs_Superlattice_Kotegawa_PRM2018}.
In contrast, the experimentally refined lattice parameters fail to show the gap opening in both systems within DFT calculations despite using different flavors of exchange correlation functional~\cite{RuPEmil2022, RuAs_Superlattice_Kotegawa_PRM2018}. 
It is indeed incredible that the relaxed structures can reproduce the gap-sizes correctly despite the well-known issue of underestimation of gap-size in plain LDA/GGA calculations, 
in sharp contrast to the DFT studies of VO$_2$, which remains metallic in the LDA/GGA and requires additional cares of electronic correlation to become an insulator~\cite{Zhu_PRB_12_VO2_mBJ,Yuan_PRB_12_VO2_U_HSE,Grau-Crespo_PRB_12_VO2_HSE,Eyert_PRL_11_VO2_HSE}.
We attribute the difference to the general expectation that the $4d$ valence electronic states in Ru\Pn~ are less localized and less magnetic than the $3d$ states in VO$_2$, making Ru\Pn~ more tractable in DFT.

The lattice parameters and energy differences for various orthorhombic and monoclinic phases obtained from  the DFT calculations are shown in 
Table S1 of Supplementary Material~\cite{Supplementary}.
For RuP and RuAs, the DFT relaxed monoclinic phases are lower in energy compared to the orthorhombic phases which is consistent with the fact that the systems prefer monoclinic distortion at low temperature. 
For RuSb, our calculation finds that the system prefers the orthorhombic phase even when the relaxation with large monoclinic distortion was attempted.
Our results from phonon calculation which is discussed below are consistent with the above findings thereby unifying our discussion.  
The DFT relaxed lattice parameters and unit-cell volume obtained using GGA exchange-correlation functional are consistently larger than the experimental values due to the well-known delocalization problem resulting in the overestimation of the lattice paramaters in GGA. 
The LDA calculations, on the other hand, show decrease in the unit-cell volume due to the well-known overestimation of the binding energy problem~\cite{DFTChallenges_CohenChemRev2012}. 
Despite the opposite trends in the structural optimization, both functionals correctly reproduces the MIT.

\emph{Electronic bands, Fermi surface and phonon dispersion.}---Fig.~\ref{fig:fig2} presents the electronic bands and Fermi surface plots for RuP and RuSb in the orthorhombic phase. Similar plots for RuAs are presented in the SM~\cite{Supplementary}. 
RuP and RuAs share similar electronic structure with degenerate flat dispersion close to the Fermi level, however the bands are less flat and slightly downshifted for RuSb.
The presence of such flat dispersion gives rise to the van Hove singularity right at the Fermi level.
The flat bands are derived mainly from  Ru-$d_{xy}$ and Ru-$d_{z^2}$ orbitals~\cite{LocalCoordinateSystem}. 
The effect of spin-orbit coupling (SOC) is small; its main role being lifting of the degeneracy of the band crossings throughout the spectrum. However, the flat band and peak DOS remains intact [see DOS comparison with and without SOC in SM]. Hence, in the following, we will present the results in the absence of the SOC. 
Note that the presence of non-symmorphic symmetry gives rise to band degeneracies and band crossing along different high symmetry directions. Some of the band crossings are topological in the sense that they are unavoidable and protected by symmetries.
Such phenomena of ``band sticking'' has been studied in detail in other systems belonging to the same space group~\cite{RhSi_Mozaffari_PRB2020,MnP_Cuono_PRM2019}.

Figs.~\ref{fig:fig2}(c) and (f) show the Fermi surface plots for RuP and RuSb, respectively.  The FS of RuP consists of a large corrugated cylinder-like electron pocket at the BZ center and small dispersionless hole FS sheets at the BZ boundary. RuAs shows similar FS topology.
However, the FS of RuSb has more structure; the electron FS at the BZ center is smaller and more elliptical, whereas the hole FS covers a much larger volume with a complicated topology.
The nesting property of the FS will be investigated in the subsequent sections.

\begin{figure*}[htb]
    \begin{center}
            \includegraphics[width=0.90\textwidth]{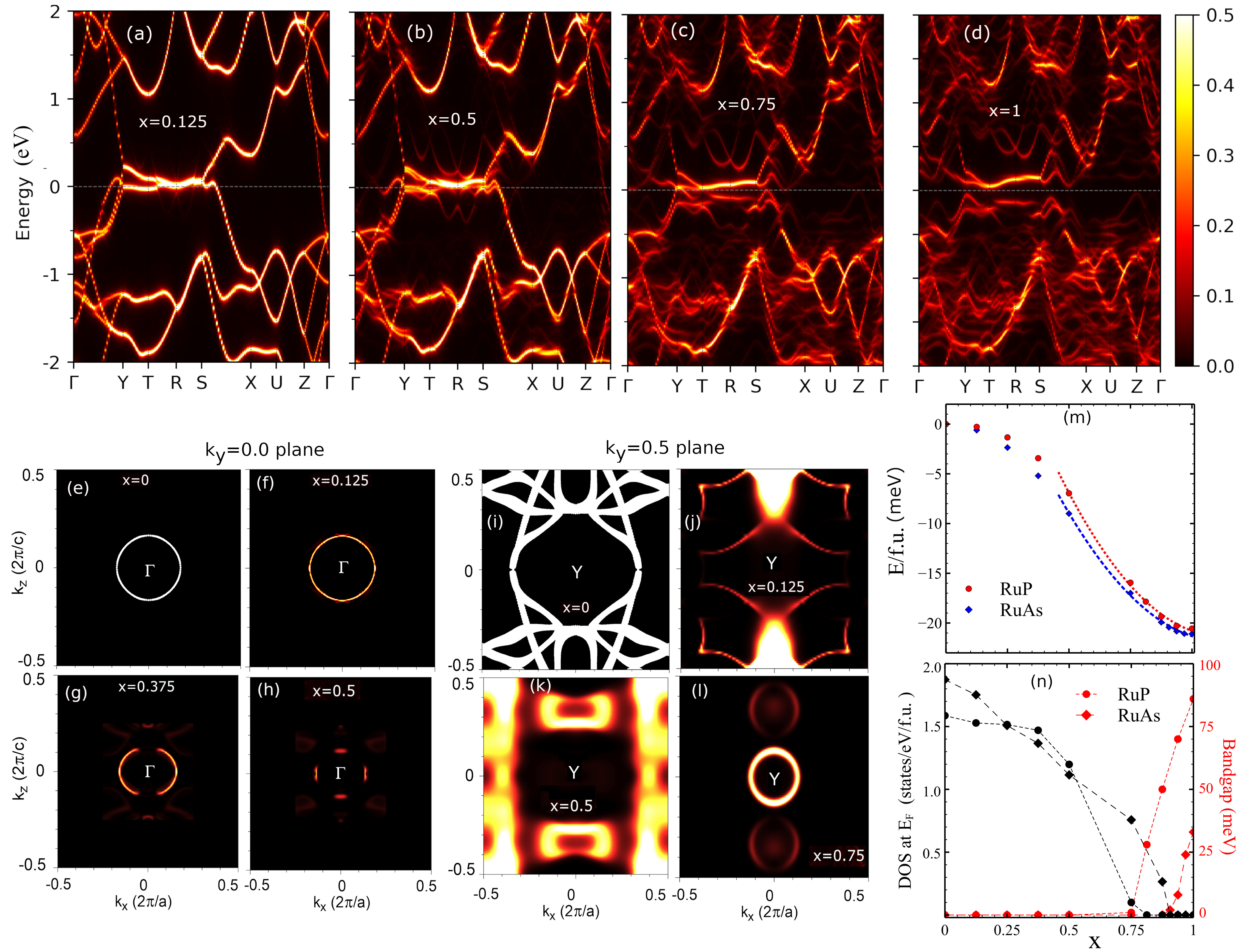}
        \end{center}
            \caption{Evolution of the electronic structure as a function of the monoclinic distortion ($x$).  (a-d) Unfolded band dispersion of the monoclinic phases of RuP. (e-h) \& (i-l) Unfolded RuP Fermi surface on the $k_x$-$k_z$ plane for $k_y=0.0$ and $0.5$ planes (in units of $2\pi/b)$, respectively. The intensity shows the spectral weights.
            (m) Total energy cost per formula unit for RuP and RuAs. The dashed lines are quadratic fits to the energy. (n) Band gap and DOS at the Fermi level. All phases, except $x$=0, have monoclinic distortion. See text for details.
    }
        \label{fig:fig4} 
\end{figure*}

Figs.~\ref{fig:fig2}(d) and (h) present the phonon dispersion for RuP and RuSb along high symmetry direction.
For RuP, there is a clear separation of the phonon modes; the low lying modes correspond to the Ru-atom displacement and high-energy modes are from the lighter P atoms. However, such distinction is less clear in RuAs and becomes even more fuzzy for RuSb owing to the larger mass of pnictogens.
Consistent with our conclusions from the lattice relaxation calculations and experimental findings, we find that orthorhombic phases of RuP and RuAs give imaginary phonon frequencies (soft modes) signalling dynamical lattice instability.
We find four soft phonon modes with Ru-displacements mainly along the zig-zag direction (\bortho).
The soft mode at T high symmetry point introduces dimerization between Ru-atoms in the zig-zag direction and is more likely to be the driver of the lattice instability.
 RuSb, on the contrary, does not show such imaginary phonon modes confirming that it's orthorhombic phase is stable at low temperature.

In order to understand more about the potential driving factors for the CDW instability, we first assess the role of the Fermi surface nesting (FSN) by calculating the real and imaginary part of the bare charge susceptibility within constant matrix approximation~\cite{Nesting_Pickett_PRB1977,FSNesting_Johannes_PRB2008} [see SM for definition].
The CDW vectors in the orthorhombic basis are ${\bf q}=(\frac{1}{3},\frac{1}{3}, 0)$ and $(0, 0, \frac{1}{3})$.
Hence, if FSN is the driving mechanism for the CDW   transition, peaks are expected in both the real and imaginary parts of the susceptibility at these {\bf q}-values~\cite{FSNesting_Johannes_PRB2006,FSNesting_Johannes_PRB2008} which correspond to the $\Gamma$-S, $\Gamma$-Z and $\Gamma$-R directions. 
However, we find that the peak position of the real and imaginary parts of charge susceptibility has no correspondence to the CDW vector which excludes FSN as the driving force for the CDW  transition (see Fig. S2 in SM~\cite{Supplementary}).
The other possible mechanism for CDW   transition is the electron-phonon coupling (EPC) effect. 
For this, we compared the relative strength of the electron-phonon coupling constant for RuP and RuSb. 
As orthorhombic phase of RuP gives imaginary phonon modes, this creates problem while calculating coupling constant. However, we were able to stabilize the orthorhombic phase by means of electronic smearing [see Fig.S4 in SM~\cite{Supplementary}]; such electronically smeared orthorhombic phases can give an estimate of the strength of the EPC parameter $\lambda$~\cite{NbS2_Wang_PRB2020,VSe2_Si_PRB2020}.
For RuP, we estimate  $\lambda$ to be $\sim$ 1.2, whereas for RuSb, we find $\lambda$ to be less than 1.
This hints that EPC could be responsible for the CDW instability in RuP and RuAs.


\begin{figure}[t]
    \begin{center}
            \includegraphics[width=0.50\textwidth]{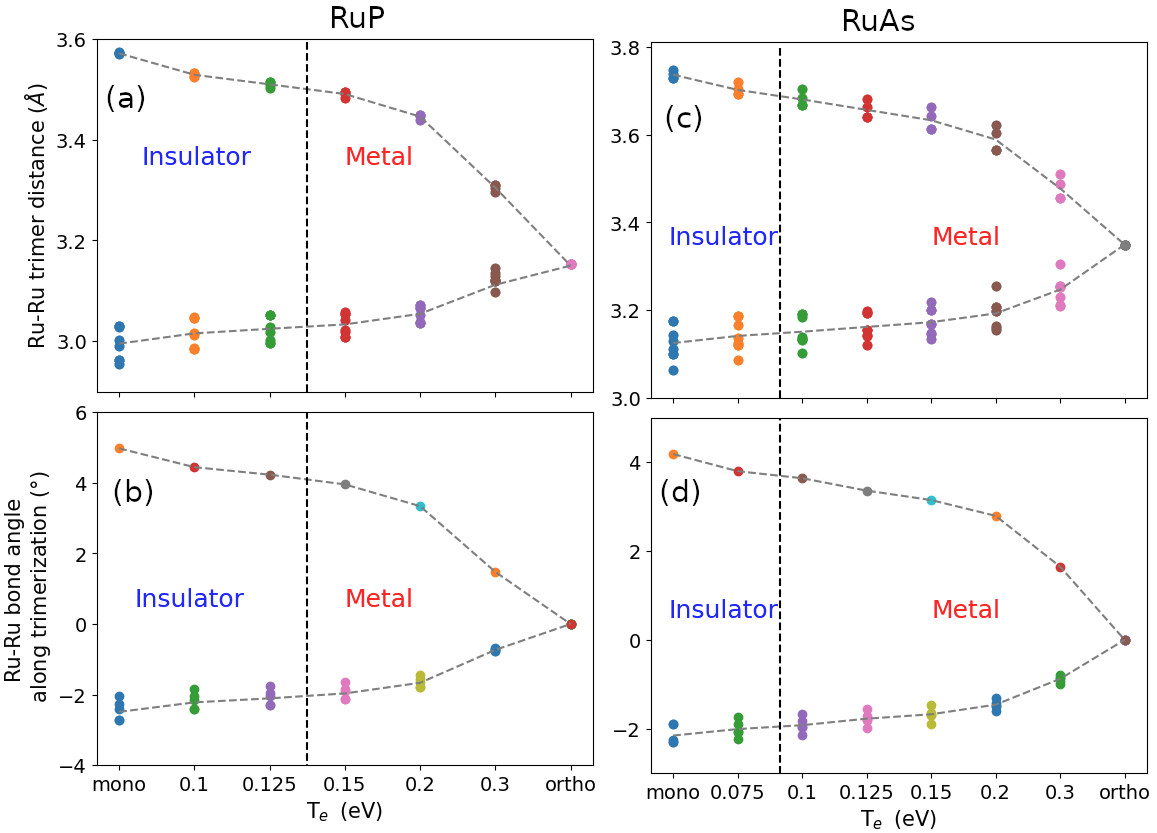}
            \end{center}
            \caption{Variation of the Ru-Ru nearest neighbour bond distance (upper panels) and bond angle (lower panels) along the trimerization direction for (a, b) RuP and (c, d) RuAs as a pngction of the electronic temperature ($T_e$). The gray dashed lines are guide to the eye. The black dashed line shows the MIT.
        }
        \label{fig:T_el} 
\end{figure}

\emph{Monoclinic distortion.}---
To explore the relationship between the monoclinic distortion and MIT, we study the evolution of the electronic structure and total energy cost as a function of the strength of the monoclinic distortion by linearly combination of the orthorhombic and monoclinic phases in the ratio of $(1-x)$ and $x$, respectively.
Figs.~\ref{fig:fig4}(a-d) and (e-l) show the unfolded band structure and Fermi surface plots, respectively, in the BZ of the orthorhombic unit cell, for various values of $x$.
The folded band structure in the BZ of the monoclinic unit cell along with the orbital character is presented in the SM~\cite{Supplementary}.
We find that the flat bands do not immediately gap-out; instead there is rather a large critical value of monoclinic distortion after which the flat bands open gap. 
The initial reaction of the system to the monoclinic distortion is to gap-out the $\Gamma$-centered electron Fermi surface [Figs.~\ref{fig:fig4}(e-h)] rather than gapping out the flat bands located on the $k_y=0.5$ plane and the flat bands undergo massive reconstruction.
At $x\approx 0.5$, the FS gaps out completely at $k_y=0$ plane but does not gap out at $k_y=0.5$ plane showing pseudo-gap like feature.
The DOS at the Fermi surface [Fig.~\ref{fig:fig4}(f)] keeps decreasing when monoclinic distortion increases with a noticeable sharp drop at the value of distortion when the flat bands gap-out.
The total energy gain as a function of $x$ [Fig.~\ref{fig:fig4}(f)] confirms that the system favors monoclinic distortion.
In addition, we find that closer to the insulating phase ($x>0.4$), quadratic equation of form $E(x) \sim -\frac{1}{2}gx + \frac{1}{2}kx^2$, where $g$ and $k$ denote the electron-lattice (\textit{e-l}) and lattice-lattice (\textit{l-l}) coupling, respectively,
 yields excellent fit to the data points, suggesting that \textit{e-l} coupling stabilizes the monoclinic phase~\cite{LaMnO3_Yin_PRL2006}. From the fits, we find the value of $g$ and $k$ to be of similar order for RuP and RuAs.

\emph{Ultrafast laser induced MIT.}---Owing to the small band-gap in the monoclinic phases of RuP and RuAs, it is plausible to induce insulator-to-metal transition (IMT) in these systems under ultra-fast laser irradiation similar to VO$_2$~\cite{Xu_NC_23_VO2_UED,Li_PRX_22_VO2_UED,Wall_Science_18_VO2_ultrafast-XRD,Morrison_Science_14_VO2_UED,Wegkamp_PRL_14_VO2_ultrafast-ARPES,Baum_Science_07_VO2_UED}.
In order to mimic the lattice response of the system to the ultrafast high energy high fluence pump, we performed DFT lattice relaxation calculations at several values of electronic temperatures ($T_{e}$) and studied the variation of the atomic coordinates as a function of $T_{e}$~\cite{Li_PRX_22_VO2_UED,Wall_Science_18_VO2_ultrafast-XRD,SnSe_ultrafast_Wei_npj2021}. We found that a nominal $T_{e}$ of $\sim0.1$~eV is sufficient to induce insulator to metal transition in RuAs whereas for RuP, the critical $T_{e}$ is $\sim0.15$~eV owing to its slightly larger band-gap.
In Fig.~\ref{fig:T_el}, we show the variation of the bond distances and bond angles along the trimerization direction as a function of $T_{e}$ for RuP and RuAs. The trimer bond distances and bond angles decrease monotonically with $T_{e}$ and finally reach the non-trimerized phase for larger values of $T_{e}$. However, the IMT happens well before the system reaches the non-trimerized phase, i.e., the monoclinic insulating phase can enter the \emph{monoclinic metallic} phase---similar to what has been reported in Ref.~\cite{Morrison_Science_14_VO2_UED} on VO$_2$---with less than 10\% decrease in the trimerization bond distance and bond angle. This is consistent with the results from linear interpolation which shows similar behavior [Fig.~3(n)]. It will be interesting to verify this prediction of two-step light-induced IMT in future ultrafast pump-probe experiments.

In summary, we perform a systematic study of the electron, phonon, and electron-lattice interplay in the parent compounds of Ru\Pn~ superconductors using first-principles calculations.
Our calculations reveal a direct correlation between the coexistence of the flat-bands, CDW instability and MIT in this family.
The electron and phonon calculations find that RuP and RuAs members of the family host flat bands at the Fermi surface and undergo monoclinic distortion at lower temperature accompanied by MIT, whereas RuSb remains orthorhombic and metallic at low temperature which is in agreement with the experimental observation.
We predict a two-step insulator to metal and structural phase transition in the monoclinic phases of RuP and RuAs which can be attested in future ultrafast experiments.
In addition, it will be interesting in the future to study the stability of the monoclinic phases of RuP and RuAs and the possible emergence of monoclinicility in RuSb by means of small external perturbations like pressure, chemical substitution, etc. 

This work was supported by the U.S. Department of Energy (DOE) the Office of Basic Energy Sciences,
Materials Sciences and Engineering Division under
Contract No. DE-SC0012704.


\begin{thebibliography}{65}
\bibitem [{Loc()}]{LocalCoordinateSystem}%
  \BibitemOpen
  \href@noop {} {}\bibinfo {note} {The local coordinate system is chosen such
  that one of the lobes of the $d_{xy}$ orbitals is oriented along the zig-zag
  chain in the \bortho~ direction and $d_{z^2}$ is along the \aortho~
  direction.}\BibitemShut {Stop}%
\end{thebibliography}

%

\end{document}